\documentclass[%
reprint,
amsmath,amssymb,
aps,longbibliography
]{revtex4-1}
\usepackage{url}
\usepackage[colorlinks=true, linkcolor=blue,urlcolor=blue,anchorcolor=blue,citecolor=blue,bookmarksnumbered]{hyperref}
\usepackage{graphicx}
\usepackage{dcolumn}
\usepackage{bm}

\begin{document}
\title{Robust beam splitter with fast quantum state transfer through a topological interface}	
\author{Jia-Ning Zhang$^{1}$}\author{Jin-Xuan Han$^{2}$}\author{Jin-Lei Wu$^{3}$}\email[]{jinlei\_wu@126.com}\author{Jie Song$^{2}$}\author{Yong-Yuan Jiang$^{1,2,4,5,6}$}\email[]{jiangyy@hit.edu.cn}
\affiliation{$^{1}$Department of Optoelectronics Science, Harbin Institute of Technology, Weihai 264209, China}
\affiliation{$^{2}$School of Physics, Harbin Institute of Technology, Harbin 150001, China}
\affiliation{$^{3}$School of Physics and Microelectronics, Zhengzhou University, Zhengzhou 450001, China}
\affiliation{$^{4}$Collaborative Innovation Center of Extreme Optics, Shanxi University, Taiyuan 030006, China}
\affiliation{$^{5}$Key Laboratory of Micro-Nano Optoelectronic Information System, Ministry of Industry and Information Technology, Harbin 150001, China}
\affiliation{$^{6}$Key Laboratory of Micro-Optics and Photonic Technology of Heilongjiang Province, Harbin Institute of Technology, Harbin 150001, China}

\begin{abstract}
The Su-Schrieffer-Heeger (SSH) model, commonly used for robust state transfers through topologically protected edge pumping, has been generalized and exploited to engineer diverse functional quantum devices. Here, we propose to realize a fast topological beam splitter based on a generalized SSH model by accelerating the quantum state transfer (QST) process essentially limited by adiabatic requirements. The scheme involves delicate orchestration of the instantaneous energy spectrum through exponential modulation of nearest neighbor coupling strengths and onsite energies, yielding a significantly accelerated beam splitting process. Due to properties of topological pumping and accelerated QST, the beam splitter exhibits strong robustness against parameter disorders and losses of system. In addition, the model demonstrates good scalability and can be extended to two-dimensional crossed-chain structures to realize a topological router with variable numbers of output ports. Our work provides practical prospects for fast and robust topological QST in feasible quantum devices in large-scale quantum information processing.

\textbf{Keywords}: quantum state transfer, beam splitter, topological router.
\end{abstract}
\maketitle

\section{Introduction}
In large-scale quantum information processing, information encoded in quantum states needs to be transmitted in a coherent manner between different nodes within a quantum network~\cite{PhysRevLett.78.3221,Matsukevich2004,PhysRevLett.92.187902,Banchi2011,PhysRevLett.108.153603,PhysRevA.102.022608}. In the last few years, great efforts have been devoted into exploring the optimal protocol for achieving efficient state transfer in the simplest and most common one-dimensional spin-1/2 chain, and the results can be further applied to incorporate various quantum systems such as quantum dots~\cite{ouyang2003,PhysRevB.70.235317,chen2012long,PhysRevLett.119.060501,Kandel2021},  coupled waveguides~\cite{PhysRevA.87.012309,chapman2016,PhysRevA.105.L061502}, superconducting circuits~\cite{PhysRevLett.100.113601,PhysRevA.98.012331,PhysRevApplied.10.054009}, and coupled-cavity arrays~\cite{PhysRevA.93.032310,PhysRevA.93.032337}. However, due to the existence of inevitable manufacturing imperfections within the devices and decoherence effect induced by the environment, the reliability of the quantum information transmission may be significantly reduced~\cite{PhysRevA.77.012303,PhysRevA.82.022336,PhysRevA.87.042311,PhysRevA.92.062305,PhysRevA.105.032612}. Therefore, it is of urgent need to improve the fidelity of quantum state transfer (QST) and circumvent the impact of different sources of disorder and decoherence.

Recently, the discovery of topological insulators opens up new prospects for efficient and robust quantum information processing~\cite{PhysRevB.31.3372,moore2010,qi2010,RevModPhys.88.035005,RevModPhys.89.041004}. Owing to their nontrivial topological energy band structures in momentum space which are inequivalent to traditional insulators, topological insulators hold simultaneously insulating bulk states and conducting edge states which can be characterized by topological invariants rooted in global geometric properties of the system~\cite{RevModPhys.82.3045,RevModPhys.83.1057,RevModPhys.88.021004,RevModPhys.91.015006}. These conducting edge states localized at the boundary of the system are inherently protected by the energy gap, and as a consequence, are immune to mild manufacturing defects or environmental perturbations and able to propagate along the boundary unidirectionally without generating back scattering~\cite{seo2010,lang2017,dlaska2017}. These prominent features make topological edge states a promising candidate for not only robust QST~\cite{PhysRevA.102.022608,PhysRevA.98.012331,lemonde2019,cao2021,cheng2022,qi2020,qi2021} but also quantum computing~\cite{stern2013,sarma2015,he2019} and quantum entanglement~\cite{PhysRevResearch.2.043191,dai2022,Han2021,Han2022}. As one of the most commonly studied one-dimensional models enabling nontrivial topological edge states, the Su-Schrieffer-Heeger (SSH) model possesses structural simplicity and rich forms of edge states. Based on the topological edge channel in the SSH lattice, the topologically protected QST between different nodes has been extensively explored and corresponding topological quantum devices has been proposed, including topological beam splitters~\cite{PhysRevA.102.022404,PhysRevB.103.085129}, topological routers~\cite{PhysRevResearch.3.023037,PhysRevApplied.18.054037}, topological lasers~\cite{st2017,PhysRevLett.120.113901,longhi2018,PhysRevApplied.13.064015,harder2021,PhysRevResearch.4.013195}, and so on. The key point of QST based on the topological edge channel is the adiabatic evolution of the channel state. The speed of QST needs to be sufficiently slow so as to avoid nonadiabatic transitions between the channel and bulk states during the whole transfer process, which results in topological quantum devices with total evolution time usually too long to be feasibly implemented in actual physical systems. Accelerated adiabatic pumping of topological edge state can be realized by designing the topological edge channel reasonably, and several protocols of QST based on the SSH model have been proposed~\cite{PhysRevB.99.155150,PhysRevResearch.2.033475,PhysRevB.102.174312,PhysRevA.103.052409}. However, these techniques, elaborately orchestrated to improve the transmission efficiency and enhance robustness against disorder and decoherence within the quantum system, focus mainly on the topologically protected state transfer from one single node to the other, and are rarely combined with specific topological quantum devices, which are not conducive to the construction of large-scale quantum networks.

In this work, we propose to realize fast and robust QST in a symmetrical topological beam splitter based on an odd-sized SSH model with alternating onsite energies and a topological interface, for which the exponential modulation of nearest neighbor coupling strengths and onsite energies account for accelerating the transfer process. The introduction of alternating onsite energies and the topological interface opens up a topological channel, through which the input state initially prepared at the interface site can be transferred to two end sites with equal probabilities and phases. We propose exponential modulation of nearest neighbor coupling strengths and onsite energies to accelerate the state transfer process whose speed is intrinsically limited by adiabatic requirements. The effect of different exponential parameters on the performance of scheme are examined, and the optimal exponential parameters for chains of different sizes are shown. Furthermore, we investigate the robustness of the topological beam splitter by taking into consideration the impact of diagonal and off-diagonal disorders and losses of system. In addition, we prove the scalability of the symmetrical beam splitter and generalize the model to two-dimensional crossed-chain structures that can be employed to implement a topological router whose number of outports can be adjusted conveniently by cross-linking different numbers of identical even-sized SSH chains via one mutual site. Finally, we stress that fast and robust QST in the proposed beam splitter and router can be realized in superconducting circuit devices under current experimental conditions, which has numerous potential applications in efficient quantum information processing and the construction of large-scale quantum networks.

\section{Physical model and engineering of topological pumping}
\subsection{Topologically protected edge states for the generalized SSH model}
\begin{figure}[t]
\includegraphics[width=1\linewidth]{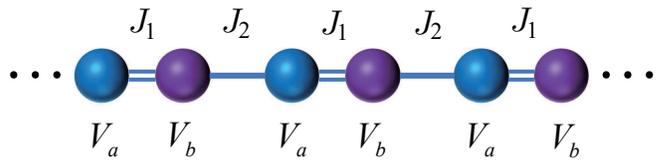}
\caption{Diagrammatic sketch of the generalized SSH model composed of $N$ unit cells. Each unit cell contains a pair of \textit{a}- (blue dot) and \textit{b}-type (purple dot) sites with onsite energies $V_a$ and $V_b$. Double lines and single lines denote the intracell coupling strength $J_1$ and intercell coupling strength $J_2$ between two adjacent sites, respectively.}\label{f1}
\end{figure}

Schematic illustration of the generalized SSH model is shown in Fig.~\ref{f1}, which describes a one-dimensional dimerized lattice composed of $N$ unit cells. The Hamiltonian of system reads as~($\hbar= 1$)
\begin{equation}\label{e1}
H=\sum_{n} V_{a}a_{n}^{\dagger}a_{n}+V_{b}b_{n}^{\dagger}b_{n}+(J_{1}a_{n}^{\dagger}b_{n}+J_{2}a_{n+1}^{\dagger}b_{n}
+{\rm H.c.}),
\end{equation}
where the first two terms represent onsite energies of two types of sites while the rest terms represent the nearest coupling between two adjacent sites. Here, $a_{n}~(a_{n}^{\dagger})$ and $b_{n}~(b_{n}^{\dagger})$ are the annihilation~(creation) operators of a particle at the $n$th \textit{a}- and \textit{b}-type sites with onsite energies $V_{a}$ and $V_{b}$, and $J_{1}$ and $J_{1}$ are the respective intracell and intercell coupling coefficients assumed to be real and positive.

For periodic boundary conditions (PBC), we can use the Bloch theorem and rewrite the bulk Hamiltonian as~
\begin{equation}\label{e2}
H_{\text{bulk}}=\sum_{\mathrm{n}=1}^{N} V_{a}a_{n}^{\dagger}a_{n}+V_{b}b_{n}^{\dagger}b_{n}+(J_{1}a_{n}^{\dagger}b_{n}+J_{2}a_{m+1}^{\dagger}b_{n}
+{\rm H.c.}),
\end{equation}
with $m=n \bmod N$. After performing a Fourier transformation $a_{n}=\frac{1}{\sqrt{N}} \sum_{k} e^{i k n} a_{k}$ and $b_{n}=\frac{1}{\sqrt{N}} \sum_{k} e^{i k n} b_{k}$,  with the wavenumber $k \in\left\{\frac{2 \pi}{N}, \frac{4 \pi}{N}, \cdots, \frac{2 N \pi}{N}\right\}$ being from the first Brillouin zone, the Hamiltonian can be moved into the momentum space, represented by $H_{\text {bulk }}=\sum_{k} V_{a} a_{k}^{\dagger} a_{k}+V_{b} b_{k}^{\dagger} b_{k}+\left[\left(J_{1}+J_{2} e^{-i k}\right) a_{k}^{\dagger} b_{k}+\text { H.c. }\right]$. For each wavenumber $k$, the bulk Hamiltonian in the momentum space under the basis of $(a_k, b_k)^T$ can be expressed as~
\begin{equation}\label{e3}
H_{k}=\left(\begin{array}{cc}
	V_{a} & J_{1}+J_{2}e^{-ik} \\
	J_{1}+J_{2}e^{ik} & V_{b}
\end{array}\right).
\end{equation}

\begin{figure}[t]
\includegraphics[width=1\linewidth]{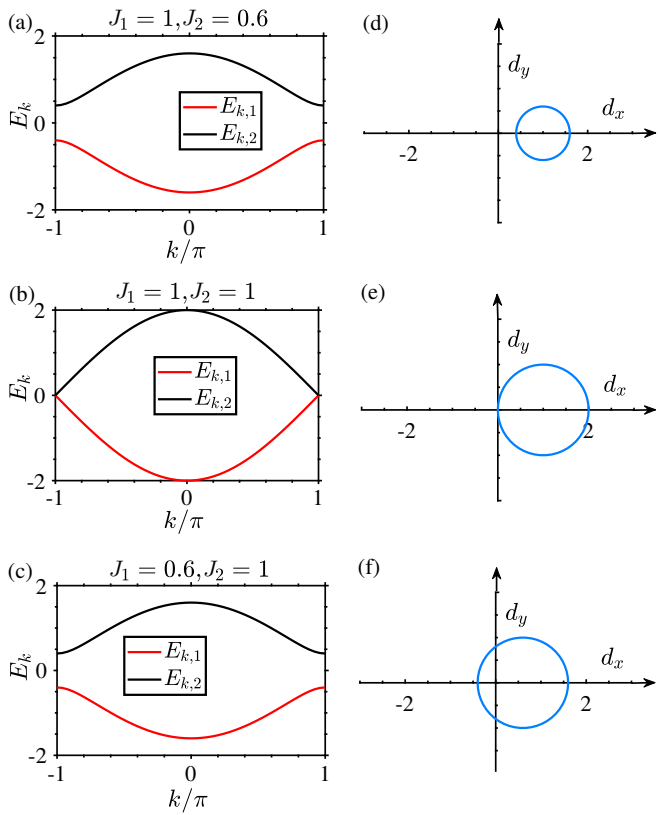}
\caption{(a)-(c)~Energy spectrum of SSH lattice in the momentum space for three settings of the coupling strengths (a) $J_1=1, J_2=0.6$; (b) $J_1=1, J_2=1$; (c) $J_1=0.6, J_2=1$. (d)-(f)~Winding of the bulk momentum-space Hamiltonian for the three settings as the wavenumber runs across the Brillouin zone.}\label{f2}
\end{figure}

\begin{figure*}
\includegraphics[width=0.98\linewidth]{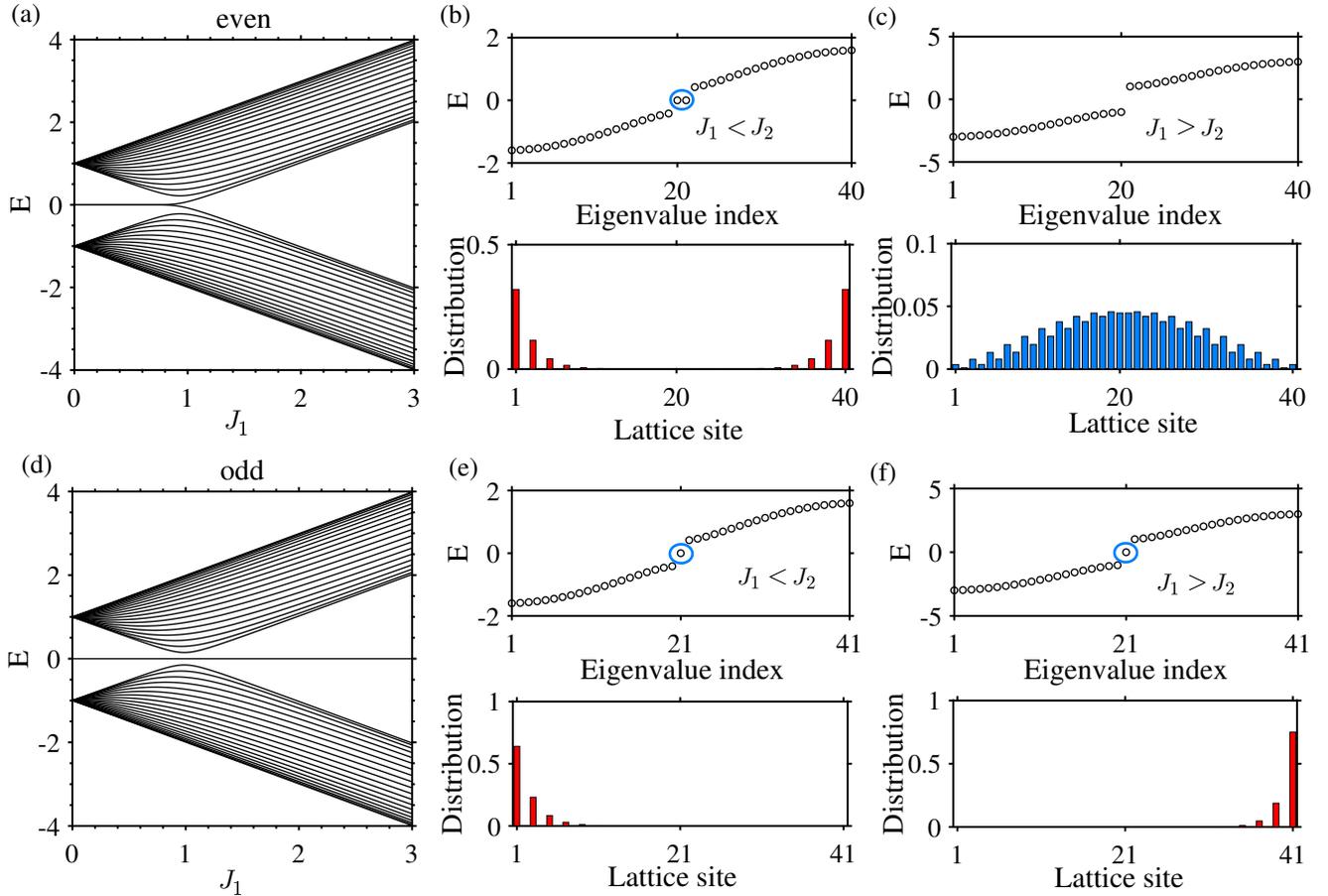}
\caption{(a)~Energy spectrum of the SSH lattice with 40 sites with varying intracell coupling $J_1$ but fixed intercell coupling $J_2=1$. $J_1<1~(J_1>1)$ corresponds to the nontrivial (trivial) topological phase. (b)~Energy spectrum (upper panel) and distribution of the gap state (lower panel) for $J_1=0.6$. (c)~Energy spectrum (upper panel) and density distribution of one bulk state (lower panel) for $J_1=2$. (d)-(f)~Energy spectra and distributions of the zero-energy edge state of the SSH lattice with 41 sites with the same coupling strengths as those in (a)-(c), respectively.}\label{f3}
\end{figure*}

We first consider the case of the standard SSH model where there is no onsite energy on the lattice sites. By diagonalizing $H_{k}$, the eigenvalues can be obtained $E_{\pm}(k)=\pm \sqrt{J_{1}^{2}+J_{2}^{2}+2 J_{1} J_{2} \cos k}$, corresponding to eigenstates~
\begin{equation}\label{e4}
|\psi_{\pm}(k)\rangle=\frac{1}{\sqrt{2}}\left[E_{\pm}(k) /\left(J_{1}+J_{2} e^{i k}\right), 1\right]^{T}.
\end{equation}
The eigenenergy spectrum of the system is divided into two bands, with an energy gap of $2\Delta$ separating the lower and filled bands from the upper and empty bands, with $\Delta=\min _{k} E(k)=\left|J_{1}-J_{2}\right|$. We plot the dispersion relation for three choices of the coupling strengths in Figs.~\ref{f2}(a)-(c). As the coupling strengths ranges from $J_1>J_2$ to $J_1<J_2$, the band gap is first closed and then reopened at the boundaries of the first Brillouin zone. Introducing the Pauli matrices $\boldsymbol{\sigma}=\left(\sigma_{x}, \sigma_{y}, \sigma_{z}\right)$ as base vectors, the Hamiltonian can be expressed in the form~
\begin{equation}\label{e5}
H(k)=\mathbf{d} \cdot \boldsymbol{\sigma},
\end{equation}
with $\mathbf{d}=\left(d_{x}, d_{y}, d_{z}\right)=\left(J_{1}+J_{2} \cos k, J_{2} \sin k, V_{a}\right)$. As displayed in Figs.~\ref{f2}(d)-(f), $\mathbf{d}$ does not enclose the origin as the wavenumber runs across the Brillouin zone for $J_1>J_2$ while encloses the origin for $J_1<J_2$, which correspond to two topological distinct phases, respectively. In the phase transition point $J_1=J_2$, eigenstates of the bulk are available with arbitrarily small energy, and the SSH model behaves like a conductor which can transport electrons from one end of the chain to the other. Otherwise, the SSH model behaves like an insulator. The encirclement of the Hamiltonian $H_{k}$ can be characterized by the winding number $w$ defined as~
\begin{equation}\label{e6}
w_{\pm}=\frac{i}{\pi} \int_{-\pi}^{\pi}\left\langle\psi_{\pm}(k) \mid \partial_{k} \mid \psi_{\pm}(k)\right\rangle.
\end{equation}
$w=0$ indicates that $H_{k}$ does not enclose the origin, corresponding to the topological trivial phase; $w=1$ indicates that $H_{k}$  encloses the origin, corresponding to the topological nontrivial phase, in which according to the body-boundary correspondence~\cite{RevModPhys.82.3045}, the SSH lattice exhibits edge states in the bulk gap under the open boundary conditions (OBC). We examine how the spectrum of an open chain changes as the intracell and intercell coupling strengths are continuously modulated. For an even-sized SSH model composed of $2N=40$ lattice sites, the topological nontrivial phase hosts two edge states exponentially localized on the boundaries of the chain, as shown in Figs.~\ref{f3}(a)-(c). By analytically solving the eigenvalue equation, we get the eigenvalues $E_{\text {even},\pm}=\pm\left|-J_{2} \frac{\left(-J_{1} / J_{2}\right)^{N}\left[\left(-J_{1} / J_{2}\right)^{2}-1\right]}{\left(-J_{1} / J_{2}\right)^{2 N}-1}\right|$, corresponding to eigenstates~
\begin{equation}\label{e7}
\left|\Psi_{\pm}\right\rangle=(|L\rangle\pm|R\rangle) / \sqrt{2},
\end{equation}
with~
\begin{equation}\label{e8}
\begin{array}{l}
	|L\rangle=|1,0,-J_{1} / J_{2}, 0, \cdots, 0,\left(-J_{1} / J_{2}\right)^{n-1}, 0, \cdots\rangle,\\
	|R\rangle=|\cdots, 0,\left(-J_{1} / J_{2}\right)^{N-n}, 0, \cdots, 0,-J_{1} / J_{2}, 0,1\rangle,
\end{array}
\end{equation}
denoting the ideal left and right edge states in thermodynamic limit~(See Appendix for more details). The degenerate gap modes of a finite-sized system in the topological nontrivial phase takes a pair of almost-zero-energy eigenvalues opposite to each other due to chiral symmetry, which makes eigenstates take the superposition of the ideal left and right edge states and localize at both ends of the system. For an odd-sized SSH model composed of $2N+1=41$ lattice sites, there is always a zero-energy edge state with eigenvector~
\begin{equation}\label{e9}
\left|\Psi_{0}\right\rangle=\left|1,0,-J_{1}/J_{2}, 0,\left(-J_{1}/J_{2}\right)^{2}, \cdots, \left(-J_{1}/J_{2}\right)^{N}\right\rangle,
\end{equation}
with the localized position depending on the ratio $J_1/J_2$, as demonstrated in Figs.~\ref{f3}(d)-(f).

\subsection{Symmetrical beam splitter via edge channel in the Rice-Mele model}
We now consider the case of the Rice-Mele model~\cite{PhysRevA.101.052323} originated from the standard SSH model by adding alternate onsite potentials $V_{a}=-V_{b}$. For each wavenumber $k$, eigenvalues and corresponding eigenstates can be obtained by analytically solving the eigenvalue equation~
\begin{equation}\label{e10}
\begin{array}{l}
	E_{\pm}(k)=\pm \sqrt{V_{a}^{2}+J_{1}^{2}+J_{2}^{2}+2 J_{1} J_{2} \cos k}, \\
	\left|\psi_{\pm}(k)\right\rangle=N_{k}\left[\left(E_{\pm}(k)+V_{a}\right) /\left(J_{1}+J_{2} e^{i k}\right), 1\right]^{T},
\end{array}
\end{equation}
with $N_{k}$ being normalization factor. In the odd-sized Rice-Mele model, there is always a gap state with eigenenergy $V_{a}$ and eigenstate $\left|\psi_{V_{a}}\right\rangle=\left|1,0,-\frac{J_{1}}{J_{2}}, 0,\left(-\frac{J_{1}}{J_{2}}\right)^{2}, 0, \cdots, 0,\left(-\frac{J_{1}}{J_{2}}\right)^{N}\right\rangle$ for which the localized position can be also modulated by tuning $J_1/J_2$, and thus can be exploited as a topologically protected quantum channel. Setting $J_1/J_2=0$ initially, $J_1/J_2=+\infty$ finally, and continuously modulating the intracell and intercell coupling strengths, a topologically protected state transfer can be realized from the left edge to the right.

\begin{figure}[b]
\includegraphics[width=1\linewidth]{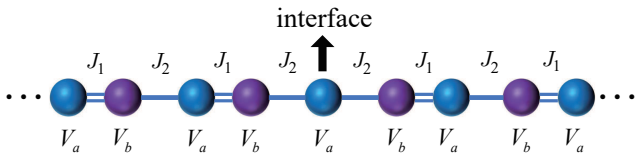}
\caption{Schematic of SSH model with the size of $L=2N+1$ with alternate onsite energies and an interface. Double and single lines denote the intracell and intercell coupling strengths $J_1$ and $J_2$ between two adjacent sites, respectively. Intracell (intercell) coupling strengths and alternate onsite energies are mirror-symmetric with respect to the interface site. $N$ is an even number, so that the topological interface falls on an \textit{a}-type site.}\label{f4}
\end{figure}

Inspired by topological edge pumping in the Rice-Mele model, a symmetrical topological beam splitter with an equal phase can be obtained based on an odd-sized SSH model with alternating onsite energies and a topological interface. We consider a finite chain comprising of $L=2N+1$ sites, with $N$ being even, i.e., structured by linking two even-sized SSH chains via one mutual \textit{a}-type site, as schematically shown in Fig.~\ref{f4}. Intracell (intercell) coupling strengths and alternate onsite energies are mirror-symmetric with respect to the interface site. The system can be described by the following interaction-picture Hamiltonian~
\begin{eqnarray}\label{e11}
H & =&\sum_{n}\left[V_{a} a_{n}^{\dagger} a_{n}+V_{b} b_{n}^{\dagger} b_{n}\right]+\left[\sum_{\mathrm{n}=1}^{N / 2}(J_{1} a_{n}^{\dagger} b_{n}+J_{2} a_{n+1}^{\dagger} b_{n})\right.\nonumber\\
&& \left.+\sum_{\mathrm{n}=N / 2+1}^{N}(J_{2} a_{n}^{\dag} b_{n}+J_{1} a_{n+1}^{\dag} b_{n})+\text { H.c.}\right].
\end{eqnarray}
In the energy spectrum of this system there always exists a gap state with eigenenergy $V_{a}$ and eigenvector~
\begin{eqnarray*}
\left|\psi_{V_{a}}\right\rangle&=&\left|1,0,-\frac{J_{1}}{J_{2}}, 0,\left(-\frac{J_{1}}{J_{2}}\right)^{2}, 0, \cdots, 0,\left(-\frac{J_{1}}{J_{2}}\right)^{N / 2}, 0, \right.\nonumber\\
&&\left. \cdots,0,\left(-\frac{J_{1}}{J_{2}}\right)^{2},0,-\frac{J_{1}}{J_{2}}, 1\right\rangle,
\end{eqnarray*}
which is localized at the topological interface when $J_1/J_2=+\infty$ but localized at both ends of the chain when $J_1/J_2=0$. Assisted by this topological edge channel, a topologically protected QST from the interface site to the two end sites with equal probabilities can be realized by continuously modulating the intracell and intercell coupling strengths from $J_1/J_2=+\infty$ to $J_1/J_2=0$. When regarding the interface site as the input port and the two end sites as two output ports, the whole system is equivalent to a symmetrical topological beam splitter, in which a particle injected into the interface site can be transferred to the two endpoints of the chain with equal probabilities. It is worth emphasizing that this beam splitting process along the gap state $\left|\psi_{V_{a}}\right\rangle$ is topologically protected by the band gap between the gap state and its adjacent bulk eigenstates and is thus immune to scattering from inherent disorders and local imperfections.

\subsection{Choice of the modulating coupling strengths and analysis of energy spectrum}\label{sec2c}
The realization of the topological beam splitter is essentially based on the adiabatic evolution of the gap state, which requires the system to be driven slowly enough so that the initial state always evolves along the gap state $\left|\psi_{V_{a}}\right\rangle$ during the transfer process. The topological pumping based on the gap state is governed by the following time-dependent Schrödinger equation~
\begin{equation}\label{e12}
i\hbar \frac{\partial}{\partial t}|\Psi(t)\rangle=H(t)|\Psi(t)\rangle
\end{equation}
where $|\Psi(t)\rangle$ can be expressed as~
\begin{equation}\label{e13}
|\Psi(t)\rangle=\sum_{n} a_{n}(t) e^{-i E_{n}(t) t}\left|\psi_{n}(t)\right\rangle,
\end{equation}
with $\left|\psi_{n}(t)\right\rangle$ and $E_{n}(t)$ obeying the instantaneous eigenequation $H(t)\left|\psi_{n}(t)\right\rangle=E_{n}(t)\left|\psi_{n}(t)\right\rangle$, and $a_{n}(t)$ denotes the probability amplitude on the $n$th instantaneous eigenstates. Substituting Eq.~(\ref{e13}) into Eq.~(\ref{e12}), we get~
\begin{equation}\label{e14}
\frac{\partial}{\partial t} a_n(t)=\sum_{m \neq n} a_m(t) e^{it[E_n(t)-E_m(t)] } 
\frac{\langle\psi_n(t)|\frac{\partial H(t)}{\partial t}| \psi_m(t)\rangle}{E_m(t)-E_n(t)}.
\end{equation}
In order to approach the adiabatic limit, we have~
\begin{equation}\label{e15}
\sum_{m \neq n} \frac{\left\langle\psi_n(t)\left|\frac{\partial H(t)}{\partial t}\right| \psi_m(t)\right\rangle}{\left|E_m(t)-E_n(t)\right|}\ll1.
\end{equation}
To satisfy the adiabatic condition, the instantaneous energy difference between the gap and bulk states should be large enough, and the derivative of the Hamiltonian, which is directly related to the slope of the driving function, should be sufficiently small. In order to enhance the speed and efficiency of state transfer, we need to adjust the intracell and intercell coupling strengths appropriately so that the system can be driven strongly where the energy gap is wide but mildly when the energy gap is narrow. 

\begin{figure}[b]
\includegraphics[width=1\linewidth]{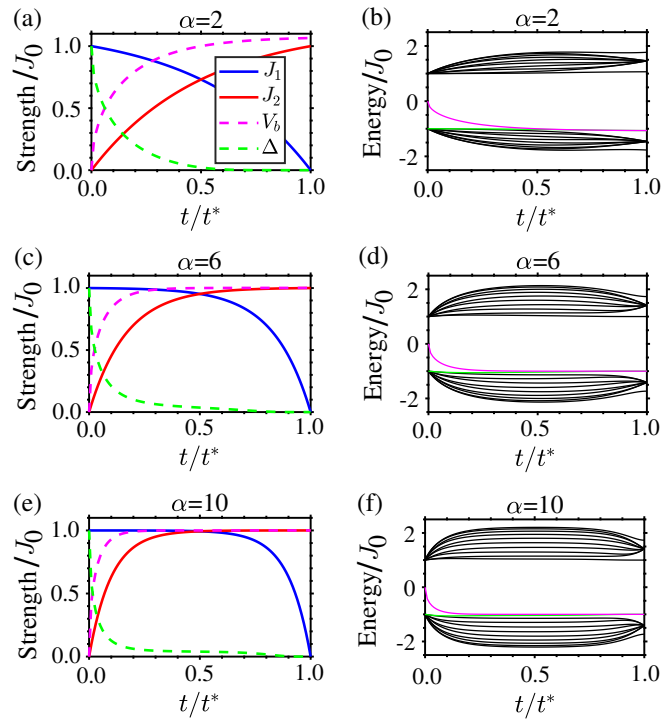}
\caption{Waveforms of coupling strengths and alternate onsite energy with (a)~$\alpha=2$, (c)~$\alpha=6$, and (e)~$\alpha=10$. (b), (d) and (f)~Instantaneous energy spectrum as a function of time for different exponential parameters in (a), (c) and (e), respectively. The total evolution time is chosen to unity and the size of chain to be 21.}\label{f5}
\end{figure}
\begin{figure}[t]
\includegraphics[width=\linewidth]{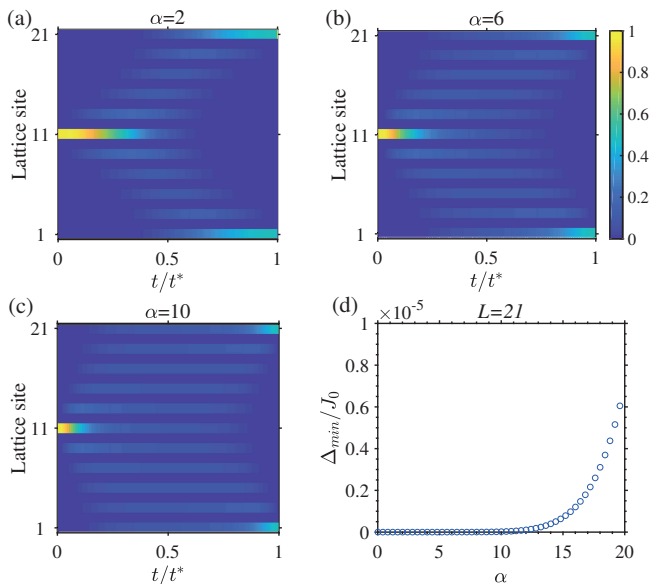}
\caption{Distribution of the gap state [magenta lines in Figs. 5(b), (d), and (f)] with eigenenergy $V_{a}$ with different values of (a)~$\alpha=2$, (b)~$\alpha=6$, and (c)~$\alpha=10$. (d)~Minimum energy gap between the gap state and the nearest-neighbor bulk states versus values of $\alpha$ with fixed chain size $L=21$.}\label{f6}
\end{figure}
Several protocols have been proposed to realize accelerated QST via topological edge channel based on the SSH model~\cite{PhysRevB.99.155150,PhysRevResearch.2.033475,PhysRevB.102.174312,PhysRevA.103.052409}. However, these techniques for accelerated adiabatic edge pumping are mainly based on the standard SSH model where there is no onsite energy on the lattice sites and focus on the topologically protected state transfer from one single node to the other. For example, Palaiodimopoulos et al.~\cite{PhysRevA.103.052409} proposed exponential modulation of nearest-neighbor coupling in an odd-sized SSH chain and achieved fast topological edge pumping. This approach, unlike the shortcuts to adiabaticity~\cite{STA2019RMP} where elaborately orchestrated counter-adiabatic terms in the Hamiltonian are introduced to suppress unwanted excitations, only involves engineering of the driving function. In this paper we adopt the exponential modulation of not only the nearest neighbor coupling strengths but also onsite energies~
\begin{subequations}
\begin{equation}\label{e16a}
	J_1=J_0 \frac{1-e^{-\alpha\left(t^\ast-t\right) / t^\ast}}{1-e^{-\alpha}},
\end{equation}
\begin{equation}\label{e16b}
	J_2=J_0 \frac{1-e^{-\alpha t / t^\ast}}{1-e^{-\alpha}},
\end{equation}
\begin{equation}\label{e16c}
	V_b=-V_a=J_0 \sqrt{\frac{J_2(2 t)}{J_0}},
\end{equation}
\end{subequations}
where $t^\ast$ denotes the total evolution time and $\alpha$ is a tunable exponential modulation parameter. According to the definitions in Eq.~(\ref{e16a}-\ref{e16c}), $J_{1,2}$ satisfy $J_1/J_2=0$ and $J_1/J_2=+\infty$ at the initial and end instants, respectively.

Taking the system with chain length of $L=21$ as an example, we select three different values of exponential parameter and plot evolution of the coupling strengths and alternating onsite energy in Figs.~\ref{f5}(a), (c), and (e). In addition, in Figs.~\ref{f5}(b), (d), and (f), we show how the instantaneous eigen-spectrum evolves over time. Referring to the gap states (magenta line) in eigen-spectrum and the corresponding evolution of onsite energy, we identify that the gap state $\left|\psi_{V_{a}}\right\rangle$ varies along the topological channel in the symmetrical beam splitting process. This can be further verified by distribution of the gap state during the evolution process as depicted in Figs.~\ref{f6}(a)-(c), in which the state transfers from the interface site to two-end sites with equal probabilities. Evolution of energy gap between the gap state and the bulk adjacent state (green line) in Figs.~\ref{f5}(b), (d), and (f) is shown in Figs.~\ref{f5}(a), (c), and (e) (green-dotted lines), respectively. In the whole process of evolution, the moment of larger (smaller) values of energy gap obviously corresponds to the larger (smaller) slope of the driving function $J_2$. Besides, as illustrated in Fig.~\ref{f6}(d), the minimum energy gap augments with the increase of parameter $\alpha$, which is positively related to the slope of the driving function. As a result, the exponential modulation of coupling strengths and alternate onsite energies is a qualified candidate to achieve fast and efficient topologically protected state transfer in the symmetrical topological beam splitter.

\begin{figure}[b]
\includegraphics[width=1\linewidth]{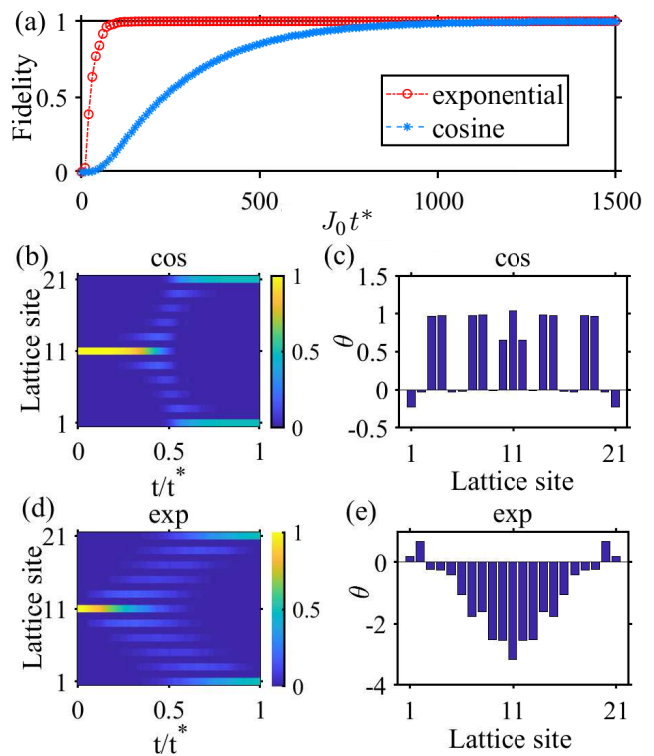}
\caption{(a~ Fidelity as a function with respect to the final time of QST for the cosine and exponential protocols. (b)-(e)~Distribution of the gap state during the evolution and the phase distribution of the evolved final state for the cosine protocol in (b) and (c), and for the exponential protocol in (d) and (e), respectively. Other parameters take $L=21$ and $\alpha=3.2$.}\label{f7}
\end{figure}

\section{Fast QST with high robustness and scalability}
\subsection{Fast QST in the topological beam splitter}
The initial state of the system~
\begin{eqnarray*}
	|\Psi_{i}\rangle&=&\left|\rho_{a,1} e^{i \phi_{a,1}}, \rho_{b,1} e^{i \phi_{b,1}},\cdots,\rho_{a,N} e^{i \phi_{a,N}}, \cdots, \right.\nonumber\\
	&& \left. \rho_{a,2N+1} e^{i \phi_{a,2N+1}}, \rho_{b,2N+1} e^{i \phi_{b,2N+1}}\right\rangle.\nonumber\\
	&=&\left|0,0,0,\cdots,0,1,0,\cdots, 0,0,0\right\rangle
\end{eqnarray*} 
is specified to set the interface site as the input port of the beam splitter. To measure how faithfully the transfer from the interface site to two end sites has occurred, we introduce fidelity defined as $F=\left|\left\langle\Psi_t \mid \Psi\left(t^\ast\right)\right\rangle\right|^2$, where~
\begin{eqnarray*}
	|\Psi_{t}\rangle&=&\left|\rho_{a,1}^{\prime} e^{i \phi_{a,1}^{\prime}}, \rho_{b,1}^{\prime} e^{i \phi_{b,1}^{\prime}},\cdots,\rho_{a,N}^{\prime} e^{i \phi_{a,N}^{\prime}}, \cdots, \right.\nonumber\\
	&& \left. \rho_{a,2N+1}^{\prime} e^{i \phi_{a,2N+1}^{\prime}}, \rho_{b,2N+1}^{\prime} e^{i \phi_{b,2N+1}^{\prime}}\right\rangle.\nonumber\\
	&=&\frac{1}{\sqrt{2}}\left|1,0,0,\cdots,0,0,0,\cdots, 0,0,1\right\rangle
\end{eqnarray*}
and $\Psi\left(t^\ast\right)$ denote the target state and the evolved state at final time $t^\ast$, respectively. Consider the system with chain size $L=21$. In order to see the speed of transfer, we plot in Fig.~\ref{f7}(a) the QST fidelity of beam splitter versus the total transfer time to compare the commonly-used cosine protocol (for example, protocols in Refs.~\cite{PhysRevA.102.022404, PhysRevB.103.085129}) and the exponential protocol with exponential parameter $\alpha=3.2$, where parameters in the cosine protocol are $J_1=\frac{J_0}{2}\left(1+\cos \frac{\pi t}{t^\ast}\right)$, $J_2=\frac{J_0}{2}\left(1-\cos \frac{\pi t}{t^\ast}\right)$, and $V_b=-V_a=J_0 \sin \frac{\pi t}{t^\ast}$. For both protocols, as total evolution time approaches infinity, the fidelity approaches unity, meaning that an excitation imposed initially at the interface site can be perfectly transferred along the chain to two-end sites with equal probabilities, which satisfies the adiabatic approximation during the transfer process, so that the system state always evolves along the gap state $\left|\psi_{V_{a}}\right\rangle$ without leaking to others. Here we suppose that the QST is successfully implemented if the fidelity is stabilized above 0.99. The implementation of the exponential protocol leads to a significantly accelerated QST process which is about 10 times faster than its cosine counterpart, since the fidelity is stabilized above 0.99 after $t^\ast=100/J_{0}$ for the exponential protocol as compared to  $t^\ast=1080/J_{0}$ for the cosine protocol. The process of QST and the phase distribution of the evolved final state for the cosine and exponential protocols are illustrated in Figs.~\ref{f7}(b)-(e), indicating that both protocols can achieve symmetrical topological beam splitting with equal phase by costing sufficient transfer time, but obviously the exponential protocol is much faster.

\begin{figure}[b]
\includegraphics[width=1\linewidth]{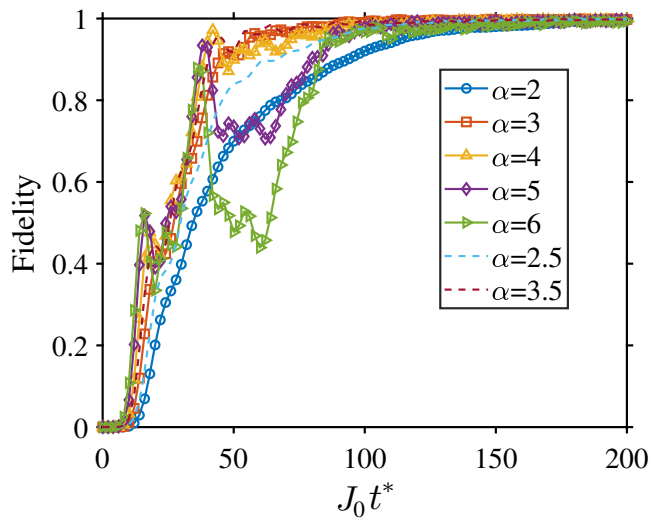}
\caption{ Fidelity as a function of the transfer time for the exponential protocol with different values of $\alpha$ with $L=21$.}\label{f8}
\end{figure}
\subsection{Effect of different values of $\alpha$}
The realization of fast QST via edge channel in the symmetrical topological beam splitter is exemplified above by setting a fine-tuned exponential parameter $\alpha=3$ in a chain of size $L=21$. We note in Sec.~\ref{sec2c} that for different $\alpha$ in the exponential modulation, there are evident differences in the slopes of the coupling functions and the corresponding energy gaps between the gap state and its nearest-neighbor bulk state in the instantaneous spectrum, leading to different effects on the QST process. To give some quantitative results,in Fig.~\ref{f8} we plot fidelity as a function of the transfer time for the exponential protocol with different $\alpha$. Taking a closer look at the fidelity curves of the exponential protocol, we notice the existence of mild oscillations, which indicates that resonant processes are at work in the QST process. When a smaller $\alpha$ is chosen, the driving function is flattened and the minimum energy gap between the gap state and its nearest-neighbor bulk state narrows down, as analytically investigated in Sec.~\ref{sec2c}, where the resonant processes are suppressed effectively and longer total transfer time is required for successful symmetrical beam splitting. Conversely, larger values of the exponential parameter $\alpha$ lead to a steeper slope of the driving function and better separation between the gap state and its nearest-neighbor bulk state. As a consequence, the resonant processes are intensified and strong oscillations appear at the fidelity curve, which takes longer time for the fidelity being stabilized at a sufficiently large value. Therefore, it is a trade-off to set $\alpha=3$ for a chain of size $L=21$, when the system is driven strongly enough to achieve high-fidelity QST of beam splitter in relatively short time, yet mildly enough to avoid strong oscillations of the fidelity curve.

\begin{figure}[t]
\includegraphics[width=1\linewidth]{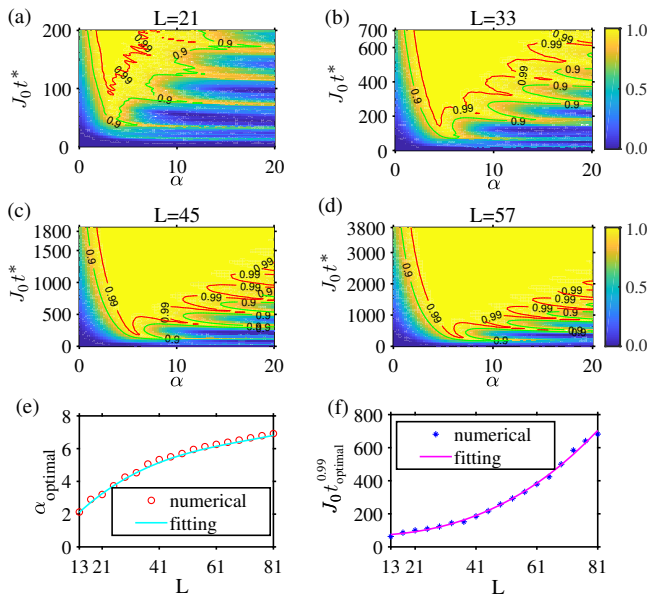}
\caption{Fidelity of QST versus varying $\alpha$ and $t^\ast$ for the exponential protocol with (a)~$L=21$, (b)~$L=33$, (c)~$L=45$, and (d)~$L=57$. The green and red solid lines represent 0.9 and 0.99 fidelity contour lines, respectively. (e)~Optimal exponential parameters and (f)~the corresponding total evolution time needed for 0.99-fidelity as a function of the size of the chain. The scattering dots and the lines represent the numerical and cubic polynomial fitting results, respectively.}\label{f9}
\end{figure}

Further, we investigate the fidelity of QST by varying $\alpha$ and the total evolution time with $L=21,~33,~45$, and $57$, respectively, as illustrated in Figs.~\ref{f9}(a)-(d). The 0.9 and 0.99 fidelity contour lines manifest intense oscillations for larger $\alpha$, yet similar to Fig.~\ref{f8}, for smaller $\alpha$ the oscillations are suppressed substantially. We can always find the optimal exponential parameter so as to reach a balance between accelerating the symmetrical beam splitting process and avoiding excessive oscillations. For instance, the optimal exponential parameters can be set as $\alpha=3.2,~4.5,~5.5$, and $6.1$ for chain sizes $L=21,~33,~45$, and $57$, respectively. As shown in Figs.~\ref{f9}(e)-(f), by selecting different chain sizes and optimizing the best exponential parameters as well as corresponding total evolution times for the final fidelity equaling to 0.99 as numerical samples, $\alpha_{\text{optimal}}$ versus $L$ and $\alpha_{\text{optimal}}^{0.99}$ versus $L$ can be fitted by cubic functions $\alpha_{\text {optimal}}=1.2 \times 10^{-5} L^3-0.0026 L^2+0.22 L-0.33$ and $J_{0}t_{\text{optimal}}^{0.99}=0.00052L^3+0.059L^2-0.34L+68$, respectively. The exponential parameter and corresponding evolution time should be large enough to satisfy the adiabatic condition for a longer size of the chain.

\subsection{Robustness against disorders and loss of system}
\begin{figure}[t]
\includegraphics[width=1\linewidth]{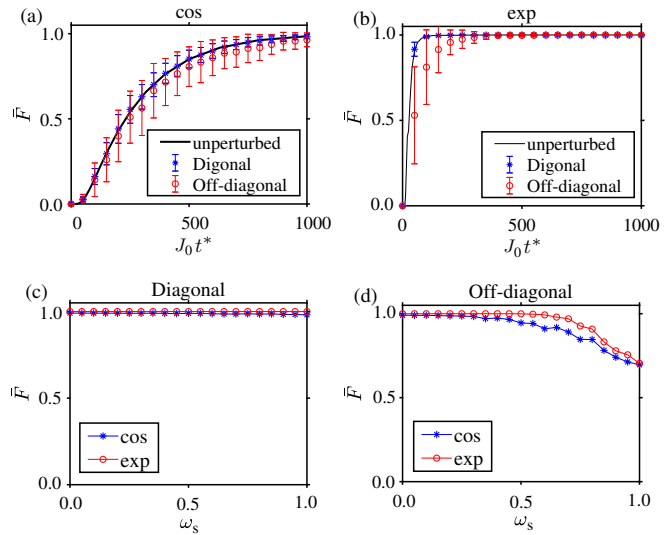}
\caption{Impact of diagonal and off-diagonal disorders with $\omega_s=0.4$ on the fidelity for the (a)~cosine and (b)~exponential protocols. Each point corresponds to the mean value of fidelity averaged over 100 disorder realizations. Average fidelity as a function of the disorder strength for (c)~diagonal and (d)~off-diagonal disorders for different protocols. Other parameters take $L=21$ and $\alpha=3.2$.}\label{f10}
\end{figure}

Due to the existence of manufacturing defects of systematic elements in practice, perfect modulation of the coupling strengths and onsite energies is almost unattainable. In this section, we examine robustness of QST in the beam splitting process by introducing disorders both in coupling strengths and onsite energies to discuss its effect on the performance of beam splitter. The disorder in coupling strengths is generally addressed as off-diagonal disorder, while the disorder in onsite energies as diagonal ones, depending on its effect on the matrix representation of the Hamiltonian. We first consider the case of symmetric distortion, in which the way each disorder realization is imposed on the system parameters can be assumed as~
\begin{eqnarray}\label{e17}
	&&J_{1(2),n}^{i} \rightarrow J_{1(2),n}^{i}\left(1+\delta J_{1(2)}^{i}\right),\nonumber\\
	&& V_{1(2),n}^{i} \rightarrow V_{1(2),n}^{i}\left(1+\delta V_{1(2)}^{i}\right),
\end{eqnarray}
where $\delta J_{1(2)}^{i}$ and $\delta V_{1(2)}^{i}$ are assumed to remain constant during the QST process, but $\delta J_{1(2)}^{i}$ and $\delta V_{1(2)}^{i}$ acquire random real values sampled from the interval $[-\omega_{s},~\omega_{s}]$, where $\omega_{s}$ is termed the disorder strength. We plot the mean fidelity of splitting QST versus total transfer time for both kinds of disorders with moderate strength $\omega_{s}=0.4$ for the cosine and exponential protocols with $L=21$ and $\alpha=3.2$ in Figs.~\ref{f10}(a) and (b), respectively. Each point corresponds to the mean value of fidelity $\bar{F}=\frac{1}{M} \sum_{i=1}^M F_i$ averaged over $M=100$ disorder realizations for the sake of universality, while the error bars correspond to the standard deviation. What we can immediately notice is that both protocols are highly robust to the diagonal disorder, because the $F$--$t^\ast$ curves almost coincide with the unperturbed curve. Besides, the transfer process is insignificantly destroyed by non-diagonal disorder applied in the both protocols. Rather, the main impact is longer total transfer time required to achieve QST of beam splitter.

In the presence of off-diagonal disorder, a longer total transfer time $t^\ast$ is required to achieve high-fidelity QST for both protocols. In Figs.~\ref{f10}(c) and (d), for different types of disorder for the cosine and exponential protocols with $\alpha=3.2$ and $L=21$, we plot the average fidelity ($M=100$) as a function of disorder strength. For the sake of comparison, we set the total transfer time $t^\ast=1100/J_{0}$ so that the QST of beam splitter can be implemented via both protocols when disorder strength equals zero. Numerical results reveal strong robustness against diagonal disorders for both protocols, because average fidelities in Fig.~\ref{f10}(c) always approach unity. However, we find that the exponential protocol manifests stronger robustness than the cosine protocol against off-diagonal disorder, as shown in Fig.~\ref{f10}(d) where the average fidelity of the former decreases significantly when $\omega_s=0.7$, while it is $\omega_s=0.4$ that the average fidelity of the cosine protocol starts to declines significantly. To sum up, the exponential protocol apparently outperforms the cosine protocol in terms of robustness to off-diagonal disorder, while both protocols are quite robust to diagonal disorder. 

\begin{figure}[t]
\includegraphics[width=1\linewidth]{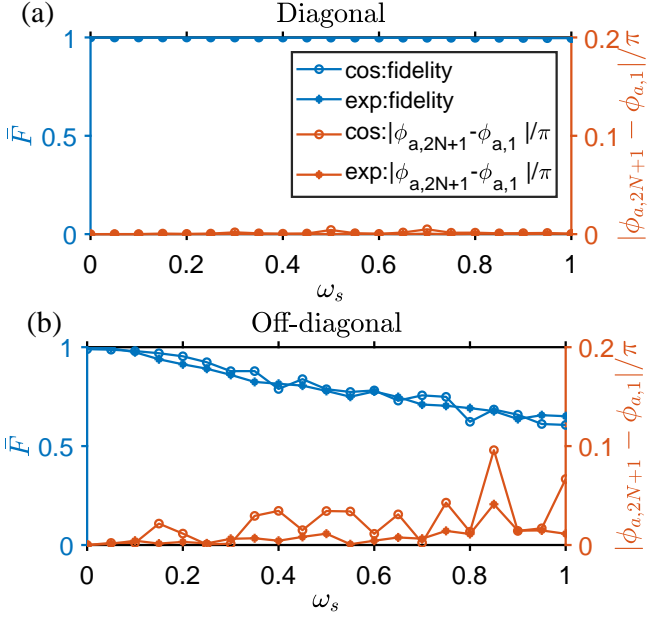}
\caption{Average fidelity and phase difference of the evolved final state at two-end sites versus disorder strength for asymmetric (a)~diagonal and (b)~off-diagonal disorders for different protocols. Other parameters take $L=21$ and $\alpha=3.2$.}\label{f18}
\end{figure}

Diagonal and off-diagonal effects as depicted in Eq.~(\ref{e17}) are mirror symmetrical with respect to the topological interface. In the following, we try to reveal the effects of asymmetric disorders. For asymmetric distortions on coupling strengths and onsite energies, their effects are described by~
\begin{equation}\label{e22}
	\begin{cases}
		J_{1,n}^{i} \rightarrow J_{1,n}^{i}, J_{2,n}^{i} \rightarrow J_{2,n}^{i}\left(1+\delta J_{1}^{i}\right),& n=1,\cdots,\frac{N}{2},\\
		J_{1,n}^{i} \rightarrow J_{1,n}^{i}, J_{2,n}^{i} \rightarrow J_{2,n}^{i}\left(1+\delta J_{2}^{i}\right),& n=\frac{N}{2}+1,\cdots,N,\\
	\end{cases}\\
\end{equation}
\begin{equation}\label{e23}
	\begin{cases}
		V_{1,n}^{i} \rightarrow V_{1,n}^{i}, V_{2,n}^{i} \rightarrow V_{2,n}^{i}\left(1+\delta V_{1}^{i}\right),& n=1,\cdots,\frac{N}{2},\\
		V_{1,n}^{i} \rightarrow V_{1,n}^{i}, V_{2,n}^{i} \rightarrow V_{2,n}^{i}\left(1+\delta V_{2}^{i}\right),& n=\frac{N}{2}+1,\cdots,N,\\
	\end{cases}\\
\end{equation}
where $\delta J_{1(2)}^{i}$ and $\delta V_{1(2)}^{i}$ acquire random real values in the interval $[-\omega_{s},~\omega_{s}]$. Taking the system of chain size $L=21$ as an example, we plot in Fig.~11 the mean fidelity of the topological beam splitter and phase difference of the evolved final state at two end sites averaged over $M=100$ samples versus disorder strength $\omega_{s}$ for the cosine and exponential protocols. We find that for both protocols, when asymmetric diagonal disorders are imposed on the system, it can still function as a robust symmetrical beam splitter with equal phase. However, in terms of robustness against asymmetric diagonal disorders, the evolved final state at the two ends may not have the same phase and amplitude, and the deviations will be amplified when $\omega_s$ increases.

\begin{figure}[t]
\includegraphics[width=1\linewidth]{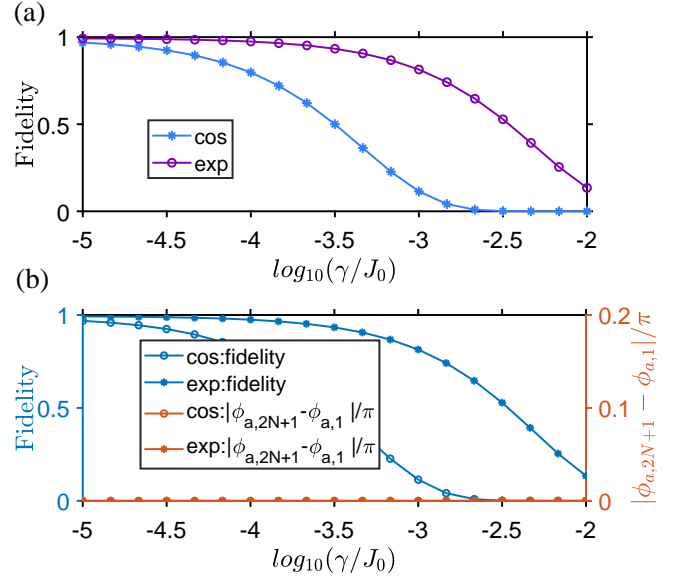}
\caption{Final fidelity as a function of the loss rate $\gamma$ for the cosine protocol with $t^\ast=1080/J_{0}$ and the exponential protocol with $t^\ast=100/J_{0}$ and $\alpha=3.2$ under symmetrical losses. (b)~Final fidelity and phase difference of the evolved final state at two-end sites versus loss rate $\gamma$ for both protocols under asymmetrical losses.}\label{f11}
\end{figure}

On the other hand, for high-efficiency QST in quantum networks, the systemic loss is an important factor on infidelity. The effect of losses during the QST of beam splitter can be considered by changing the Hamiltonian to a non-Hermitian form
\begin{equation}\label{e18}
H'=H-i \sum_n\left[\gamma_n^a a_n^{\dagger} a_n+\gamma_n^b b_n^{\dagger} b_n\right],
\end{equation}
where $H$ is the lossless Hamiltonian, and $\gamma_n^{a,b}$ denotes the loss rate of each type of sites. We first consider the impact of symmetrical loss. For convenience, we assume $\gamma_n^a=\gamma_n^b=\gamma$. The dynamics of system is governed by the non-Hermitian Lioville equation $\dot{\rho}=-i\left(H' \rho-\rho H'^{\dagger}\right)$. We plot in Fig.~\ref{f11}(a) effects of symmetrical loss on the final fidelity of QST for a chain of size $L=21$ for the cosine and exponential protocols with $\alpha=3.2$  and  transfer time fixed at $t^\ast=1080/J_{0}$ and $t^\ast=100/J_{0}$, respectively, so that the QST can be successfully implemented via both protocols when no loss exists. Evidently, the exponential protocol manifests a notable improvement of the final fidelity compared to the cosine protocol. The weakened damage of the loss effect to the QST process can be attributed to the shorter accumulation time of decoherence during the QST in the exponential protocol. We also demonstrate in Fig.~\ref{f11}(b) effects of asymmetrical loss on the final fidelity and phase difference of the evolved final state at two end sites. Here, we assume~
\begin{eqnarray*}
	&&\gamma_n^a=\gamma,\nonumber\\
	&& \gamma_n^b=\gamma(1+\delta \gamma_{1}),\quad n=1,\cdots,\frac{N}{2},\nonumber\\
	&& \gamma_n^b=\gamma(1+\delta \gamma_{2}),\quad n=\frac{N}{2}+1,\cdots,N,
\end{eqnarray*}
where $\delta\gamma_{1(2)}$ acquire random real values in the interval $[-0.1,~0.1]$. Apparently, numerical results for effect of asymmetrical loss have few differences from its symmetrical counterpart.

\subsection{Scalability}
\subsubsection{Size of chains}
\begin{figure}[t]
\includegraphics[width=1\linewidth]{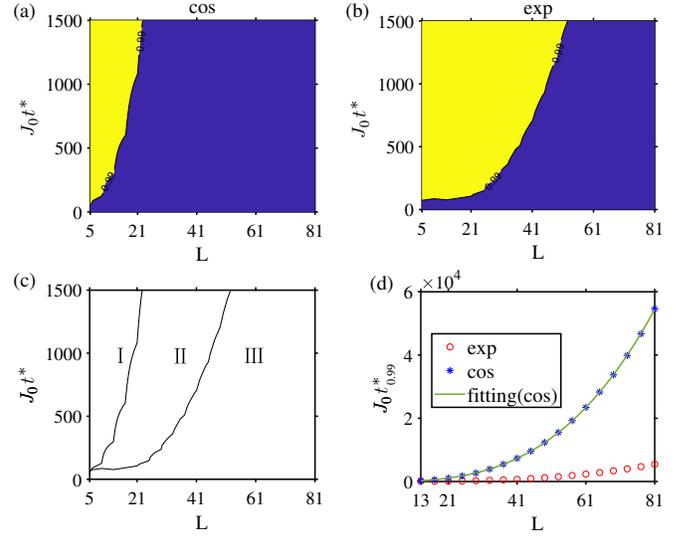}
\caption{Phase diagram of the QST in the parameter space $(t^\ast, L)$ of the (a)~cosine and (b)~exponential protocols. (c)~The total phase diagram derived from (a) and (b). (d)~The transfer time $t_{0.99}^\ast$ that each protocol takes to reach 0.99 fidelity as a function of the size of the system. $\alpha=3.2$ is used here.}\label{f12}
\end{figure}

As noted above, the exponential protocol is a compelling alternative for fast and robust QST in the symmetrical topological beam splitter, which exhibits a remarkable improvement in speed of beam splitting and robustness against both types of disorders compared to the commonly-used cosine protocol. In order to verify more extensively the effect of exponential coupling strengths and onsite energies, one crucial aspect determining the efficiency of QST is its scalability. In the following, we focus on how the exponential protocol behaves when the system size is altered. We show the phase diagram of the QST in the parameter space $(t^\ast,L)$ of the cosine and exponential protocols in Figs.~\ref{f12}(a) and (b), respectively, with the exponential parameter fixed at $\alpha=3.2$. The yellow~(purple) areas indicate that the QST of beam splitter can be implemented with fidelity over~(below) 0.99. Evidently, the parameter space can be divided into three regions according to different phase boundaries, as demonstrated in Fig.~\ref{f12}(c). In region \uppercase\expandafter{\romannumeral1}, the symmetrical beam splitter can be faithfully realized via both protocols. In region \uppercase\expandafter{\romannumeral2}, beam splitting is faithfully implemented via only the exponential protocol, while in region \uppercase\expandafter{\romannumeral3} neither protocols cannot work well. Consequently, we can choose feasible modulation protocols according to different parameter design in the system. We plot in Fig.~\ref{f12}(d) the total transfer time $t_{0.99}^\ast$ each protocol takes to achieve beam splitting as a function of the size of the system, where $t_{\text{cos}}^{0.99}$ versus $L$ for the cosine protocol can be fitted by cubic function $J_{0}t_{\text{cos}}^{0.99}=0.1L^3-0.46L^2+28L-260$. Obviously, the symmetrical topological beam splitter modulated by both protocols needs a longer total evolution time with the augmentation of chain size $L$. Nevertheless, it is evident that the exponential protocol outperforms the cosine protocol in terms of the transfer speed and manifests good scalability within the range of length we have considered here.

\begin{figure}[b]
\includegraphics[width=1\linewidth]{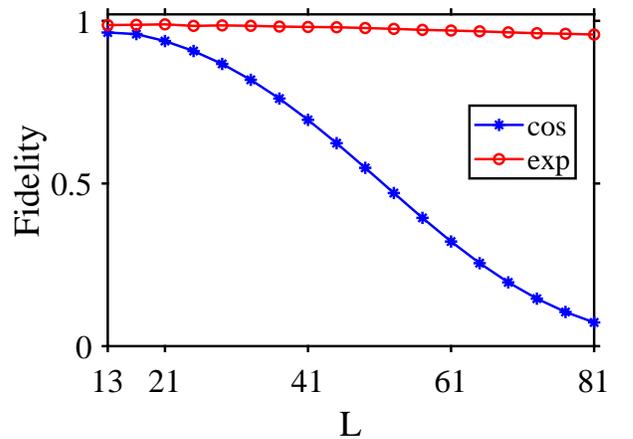}
\caption{Fidelity as a function of the chain size with fixed loss parameter $\gamma=2.5\times10^{-5}J_{0}$ for both protocols. $\alpha=3.2$ is used here.}\label{f13}
\end{figure}

In the following, we take into consideration the inevitable loss of the system. As shown in Fig.~\ref{f13}, we plot for both protocols the final fidelity as a function of the chain size with fixed loss parameter $\gamma=2.5\times10^{-5}J_{0}$. The parameter selection is made in accordance with current technological capabilities in the superconducting circuit devices, where the initial~(end) coupling strengths between resonators and the decoherence rates of photon in superconducting resonator are set to be $J_{0}/2\pi=100\rm{MHz}$ and $\gamma/2\pi=2.5\rm{kHz}$, respectively~\cite{Schmidt2013,Frunzio2011,PhysRevA.86.023837}. The exponential parameters are chosen as $\alpha=1.2 \times 10^{-5} L^3-0.0026 L^2+0.22 L-0.33$ for chains of different sizes, and the total evolution times are set to be $J_{0}t=0.1L^3-0.46L^2+28L-260$ for the cosine protocol and $J_{0}t=0.00052L^3+0.059L^2-0.34L+68$ for the exponential protocol, which ensures the 0.99-fidelity QST for both protocols in the absence of losses. The results indicate that the exponential protocol not only exhibits stronger robustness against environment-induced decoherence, but also manifests good scalability, whereas the fidelity of the cosine protocol plummets with increasing size.

\begin{figure}[t]
\includegraphics[width=1\linewidth]{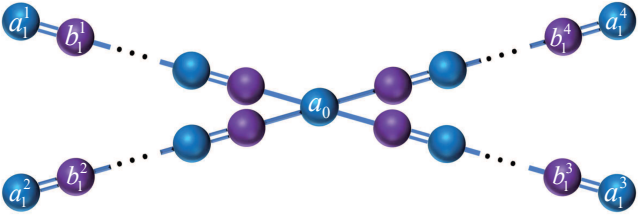}
\caption{Schematic illustration of the crossed-chain structure comprised of four even-sized SSH chains connected to a mutual additional \textit{a}-type site.}\label{f14}
\end{figure}

\subsubsection{Number of chains}
\begin{figure}[b]
	\includegraphics[width=1\linewidth]{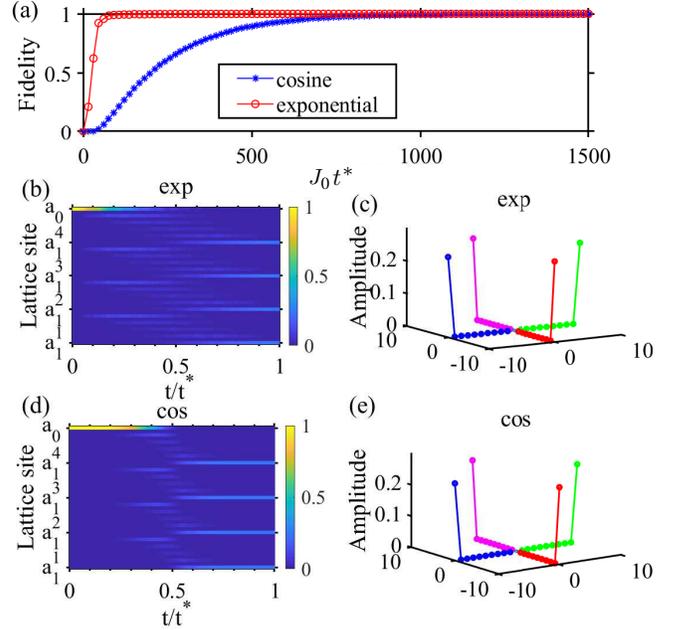}
	\caption{(a)~Final fidelity of four-outport router as a function of the transfer time for the cosine and exponential protocols. (b)-(e)~Distribution of the gap state with energy eigenvalue of $V_{a}$ during the evolution and amplitude distribution of the evolved final state for the exponential protocol in (b) and (c), and the cosine protocol in (d) and (e), respectively. Other parameters take $L=4N+1=41$ and $\alpha=3.2$.}\label{f15}
\end{figure}

The symmetrical topological beam splitter based on an odd-sized SSH model with alternating onsite energies and a topological interface can be regarded structurally as a system composed of two even-sized SSH chains connected to a mutual additional \textit{a}-type site. Another vital direction for scalability is how the exponential protocol behaves when the number of constituent chains in the crossed-chain structure is altered. To this end, in Fig.~\ref{f14} without loss of generality we consider a crossed-chain structure comprised of $K=4$ even-sized SSH chains connected to a mutual additional \textit{a}-type site, which will be taken as a typical example and analyzed extensively in the following. The interaction of the crossed-chain structure formed by $L=4N+1$ sites can be described by the following interaction-picture Hamiltonian
\begin{eqnarray}\label{e19}
H^{\prime}&=&\sum_{\sigma} \sum_{n}\left[V_{a} a_{n}^{\sigma \dagger} a_{n}^{\sigma}+V_{b} b_{n}^{\sigma \dagger} b_{n}^{\sigma}\right]+\sum_{\sigma} \sum_{\mathrm{n}=1}^{N / 2}\left[J_{1} a_{n}^{\sigma \dagger} b_{n}^{\sigma}\right.\nonumber\\
&&\left.+J_{2} a_{n+1}^{\sigma \dagger} b_{n}^{\sigma}+\text { H.c. }\right],
\end{eqnarray}
where $a_{N/2+1}^{\sigma}=a_{N/2+1}=a_{0}$, $a_{n}^{\sigma}$ and $b_{n}^{\sigma}$ are the amplitudes at the $n$th \textit{a}- and \textit{b}-type sites in a single SSH chain indexed by $\sigma$. If we regard the connecting site as the input port and the $K$ end sites as $K$ output ports, the crossed-chain structure is equivalent to a \textit{topological router}, in which a particle injected into the connecting site can be transferred to $K$ end sites with equal probabilities. To verify the fast QST for the exponential protocol in topological router with four output ports, we plot the fidelity versus the total transfer time for the cosine and exponential protocols with $\alpha=3.2$ in Fig.~\ref{f15}(a). The QST process for the exponential protocol is still about 10 times faster than its cosine counterpart, since fidelity is stabilized above 0.99 after $t^\ast=91/J_{0}$ for the exponential protocol as compared to $t^\ast=935/J_{0}$ for the cosine protocol. The process of QST and the amplitude distribution of the evolved final state under the basis of 
\begin{eqnarray*}
C&=&\left(a_{1}^{1}, b_{1}^{1}, \ldots, a_{N / 2}^{1}, b_{N / 2}^{1}, a_{1}^{2}, \ldots, b_{N / 2}^{2}, a_{1}^{3}, \ldots, b_{N / 2}^{3}, a_{1}^{4},\right. \nonumber\\
&&\left.\ldots, b_{N / 2}^{4}, a_{N / 2+1}\right)
\end{eqnarray*}
for the exponential and cosine protocols are illustrated in Figs.~\ref{f15}(b)-(e), implying that both protocols can achieve successful topological routing under sufficient transfer time, but the former protocol is obviously faster.

\begin{figure}[b]
	\includegraphics[width=0.87\linewidth]{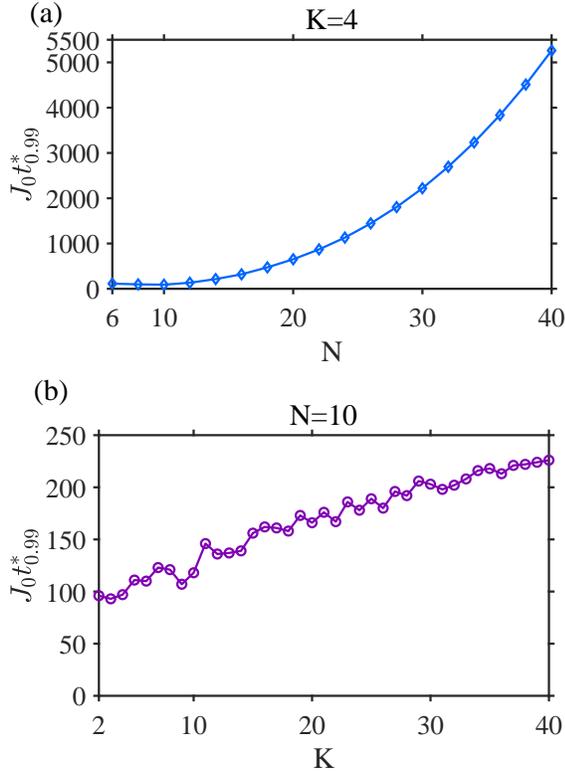}
	\caption{Using the exponential protocol with $\alpha=3.2$ for the four-outport router, transfer time $t_{0.99}^\ast$ as a function of (a)~the size of each constituent chain with $K=4$ and (b)~the number of constituent chains with $N=10$.}\label{f16}
\end{figure}

We plot in Fig.~\ref{f16}(a) the transfer time $t_{0.99}^\ast$ as a function of the size $N$ of each constituent chain. Total transfer time increases with the augmentation of the size of each constituent chain for the four-chain structure, which is consistent with the results in the two-chain beam splitter. The transfer time $t_{0.99}^\ast$ as a function of the number of constituent chains is illustrated in Fig.~\ref{f16}(b), where exponential parameter is set as $\alpha=3.2$ and the size of each constituent chain is chosen to be $N=10$. We can see that in general, total transfer time increases with the augmentation of the number of constituent chains connected in the crossed-chain structure, with mild fluctuations which can be attributed to inevitable oscillation in the $F$--$t^\ast$ curves. Therefore, by modulating the number of crossed linked chains, the number of outports can be adjusted conveniently. Such good flexibility and scalability give the topological beam splitter and topological router broad application prospects in quantum information distribution and large-scale quantum information network construction.

\begin{figure}[b]\centering
\centering
\includegraphics[width=1\linewidth]{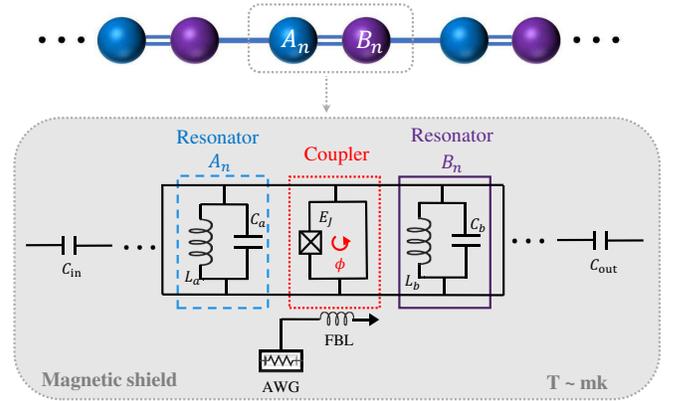}
\caption{Equivalent circuit of the coupled superconducting resonator system. Circuit elements are used to model the microwave resonator $A_n$~($B_n$) and the coupler with the additional Josephson junction in a dilution refrigerator (with a temperature $T \sim$ mK), which is placed in a magnetic shield.  The microwave resonator $A_n$~($B_n$) is an $LC$ circuit composed of a spiral inductor $L_a$~($L_b$) and a capacitor $C_a$~($C_b$). The external classical filed can be attained independently via changing the magnetic flux $\phi$ threading on the loop of coupler, which can add the FBL to connect with an AWG by adopting controlled voltage pulses~\cite{Peropadre2010,DiCarlo2009}.}\label{f17}
\end{figure}

\section{Experimental Consideration and Anticipated Improvement}
\subsection{Experiment consideration for superconducting circuit devices}
This protocol for realizing fast and robust QST in a symmetrical topological beam splitter is applicable to superconducting circuit devices, which benefits from existing circuit-QED technologies.

We can construct a superconducting resonator chain to arrange alternately the resonator $A_n$ and the resonator $B_n$ in one-dimensional space, whose equivalent circuit of one unit cell is shown in Fig.~\ref{f17}. The resonator $A_n$~($B_n$) is composed by a spiral inductor $L_a$~($L_b$) and a capacitor $C_a$~($C_b$) in analogy with harmonic oscillator, which has a single mode. In terms of the capacitor charge $Q_a$~($Q_b$) and the inductor current $I_a$~($I_b$), the Hamiltonian of oscillator is written as
\begin{eqnarray}\label{e20}
\hat{H}_{LC}&=&\frac{Q_{j}^2}{2C_{j}}+\frac{\Phi_{j}^2}{2L_{j}}
\end{eqnarray}
where $\Phi_j$ is the flux through the inductor $L_j$, and $Q_j$ is the charge on the capacitor $C_j$~($j=a,b$).  Based on the standard quantization process of an $LC$ circuit~\cite{GU2017}, the Hamiltonian of the resonator $A_n$~($B_n$) can be further written as $H_{\rm{LC}}= \hbar V_{j}j^{\dagger}j$  in terms of the creation and annihilation operators defined by $j^{\dagger}=1/\sqrt{2 \hbar V_{j}}(Q_j/\sqrt{C_j}-i\Phi_{j}/\sqrt{L_j})$ and
$j=1/\sqrt{2 \hbar V_{j}}(Q_j/\sqrt{C_j}+i\Phi_{j}/\sqrt{L_j})$, where $V_{j}=1/\sqrt{L_jC_j}$ is the oscillator frequency. Thus, onsite energies of two types of sites $A_n$~($B_n$) can be engineered in a large range of possible values by adjusting the parameters $L_j$ and $C_j$. In experiment, the onsite energies of the resonator can be selectively controlled by a DC bias voltage supply connected with the variable capacitor via a low-pass filter. The dependence of its resonant frequency on the DC bias is observed with no hysteresis, which is of great value for tunability~\cite{He_2019}.

Furthermore, a direct tunable coupler is realized by a tunable circuit element between the resonators, e.g., a flux-biased direct-current superconducting quantum interference device (SQUID)  to generate strong resonant and nonresonant tunable interactions between any two lumped-element resonators. In this work, we adopt a direct tunable coupler replaced by a SQUID between resonator $A_n$ and $B_n$, which is composed by the additional Josephson junction $E_J$~\cite{Altomare2010,Allman2014}. The flux $\phi$ threading the SQUID loop gives rise to a circulating current $I_s (\phi)=-I_c\sin(2\pi\phi/\phi_0$), where $\phi_0$ is the flux quantum. Here, $\phi$ is the sum of the externally applied flux $\phi_{\rm{ext}}$ and the flux generated by $I_s$, which can be represented by $\phi=\phi_{\rm{ext}}+L_sI_s(\phi)$. Thus, the flux dependent coupling between resonators is written as $J_{1,2}=-\sqrt{\frac{V_a}{L_a}} \sqrt{\frac{V_b}{L_b}}\frac{L^2_0}{L(\phi)}$, where $L_0$ is the inductance of the segment shared between resonator and SQUID and the effective SQUID inductance with respect to external fluxes	$L(\phi)=\partial \phi_{\rm{ext}}/\partial I_s$~\cite{Wulschner2016}. Therefore, the simplest way to tune the coupling strength $J_{1,2}$ is to apply a control magnetic flux to this loop dynamically with $\phi_{\rm{ext}}(\vec{x})= \int_{s} \textbf{B}(\vec{x},t)\cdot d\textbf{S}$, by adding the external flux-bias line~(FBL) to connect with an arbitrary waveforms generator (AWG) by adopting controlled voltage pulses~\cite{Peropadre2010,DiCarlo2009}, as shown in Fig.~\ref{f17}. Thus, superconducting circuits possessing advantages of flexibility, scalability and tunability~\cite{Blais2004,Clarke2008,You2011}, providing an excellent platform for realizing fast and robust QST in a symmetrical topological beam splitter with high fidelity.

\subsection{Possibility of further accelerating QST process in the symmetrical beam splitter}

As noted above, we have realized fast QST in the symmetrical beam splitter through exponential modulation of the nearest-neighbor coupling strengths and the onsite energies. The scheme acclerates the beam splitting process through subtle control of the driving functions according to the instantaneous eigenspetrum and is still limited by the adiabatic requirements. To further accelerate the QST process, we can consider incorporating moderately nonadiabatic resonant process between eigenstates into the adiabatic process. For example, a fast topological edge pumping protocol in which quantum state transfers rapidly from the left edge to the right was recently presented in an SSH chain~\cite{PhysRevB.102.174312}.
The intracell and intercell nearest neighbor coupling strenghts are governed by the following 3-step modulation functions~
\begin{subequations}
	\begin{equation}\label{e21a}
		J_{1}=\begin{cases}
			J_{1}(0), & t \leq t^{*}-t_{op} \\
			\frac{(J_{1}(0)-J_{2}(0))t^{*}}{t_{op}}\left(1-\frac{t}{t^{*}}\right), & t>t^{*}-t_{op} \\
		\end{cases}
	\end{equation}
	\begin{equation}\label{e21b}
		J_{2}=\begin{cases}
			J_{2}(0)+\frac{(J_{1}(0)-J_{2}(0))t}{t_{op}}, & t \leq t_{op} \\
			J_{1}(0), & t>t_{op} \\
		\end{cases} \\
	\end{equation}
\end{subequations}
where $J_{1}(0)$ and $J_{2}(0)$ denote the intracell and intercell coupling coefficients at the initial moment, $t^{*}$ is the total evolution time, and $t_{op}$ is the time interval for the coupling strengths to increase (decrease) from initial to terminal value (and vice versa). $J_{1}$ and $J_{2}$ are of mirror symmetry, so $t_{op}$ is the only free parameter. It is possible to find the best $t_{op}$ value through parameter optimization so as to produce the fastest topological egde pumping.

Different from the case of the commonly-used trigonometric protocol in which eigenenergy of the edge mode for an odd-sized SSH model remains constant during the transfer process, for the 3-step protocol the instantaneous eigenenergy are bended and its mean value is significantly increased. Therefore, the timescales which may be considered inversely proportional to the time average of eigenenergy of the edge state effectively decrease. The key to this scheme is incorporating the nonadiabatic resonance transition, whose dynamical evolution is closely related to the pulse area. Analogously, with the help of such a 3-step modulation protocol, it is of great potential to furthur speed up the QST in the symmetrical beam splitter proposed here.

\section{Conclusion}
To sum up, we have proposed a protocol of fast and robust topological pumping via edge channel through exponential modulation of the driving functions for generating a symmetrical topological beam splitter and further for deriving a topological router. We show both analytically and numerically that by continuously modulating the intracell and intercell coupling strengths and onsite energies, we can achieve topologically protected quantum state transfer (QST) from the interface site towards end sites with equal probabilities. Based on numerical analysis of the instantaneous energy spectrum of the system, we confirm that the value of the instantaneous energy gap suitably adapts to the slope of the driving functions, and then present numerical evidence of accelerated adiabatic edge pumping in the symmetrical beam splitter. Furthermore, we investigate how the selection of the exponential parameter impact the QST process. The robustness of the topological beam splitter is extensively discussed by taking into consideration the impact of diagonal and off-diagonal disorders and systematic losses. In addition, we prove the scalability in the size and number of chains for the symmetrical beam splitter and topological router, respectively.  Last but not least, we propose superconducting circuit devices as a feasible platform to implement fast and robust QST in the symmetrical beam splitter discussed in this article. The scheme provides detailed assumption of topological beam splitter and topological router assisted by fast and robust topological edge pumping, which is expected to make substantial contribution to efficient quantum information processing and the construction of large-scale quantum networks.\\

\section*{Acknowledgements}
The authors acknowledge the financial support by the National Natural Science Foundation of China (Grant No. 62075048) and Natural Science Foundation of Shandong Province of China (Grant No. ZR2020MF129).

\section*{Appendix: Hybridized edge states for even-sized SSH model}
\setcounter{equation}{0}
\renewcommand\theequation{A\arabic{equation}} 
For the even-sized SSH model, the matrix representation of the Hamiltonian on a real-space basis reads~
\begin{equation}
	H_{M}=\left(\begin{array}{cccccccc}
		0 & J_{1} & 0 & 0 & \cdots & 0 & 0 & 0 \\
		J_{1} & 0 & J_{2} & 0 & \cdots & 0 & 0 & 0 \\
		0 & J_{2} & 0 & J_{1} & \cdots & 0 & 0 & 0 \\
		\vdots & \vdots & \vdots & \vdots & \vdots & \vdots & \vdots & \vdots \\
		0 & 0 & 0 & 0 & \cdots & J_{2} & 0 & J_{1} \\
		0 & 0 & 0 & 0 & \cdots & 0 & J_{1} & 0
	\end{array}\right).
\end{equation}
In the following we prove that in the thermodynamic limit of $N \to \infty$, the topological nontrivial phase hosts two edge states localized on the boundaries of the chain. We first consider a semi-infinite long SSH lattice with left boundry. In topological nontrivial phase, we assume that the SSH lattice exhibits zero-energy eigenstate~
\begin{equation}
	|L\rangle=|\psi_{a_{1}},\psi_{b_{1}},\psi_{a_{2}}, \psi_{b_{2}}, \cdots, \psi_{a_{n}},\psi_{b_{n}}, \cdots\rangle,
\end{equation}
where $\psi_{a_{n}}~(\psi_{b_{n}})$ are the amplitudes on lattice site  $a_{n}~(b_{n})$. Solving the eigenvalue equation $H_{M}|L\rangle=0$~($H_{M}$ under right semi-infinite boundary condition), we get~
\begin{subequations}
	\begin{equation}
		J_{1}\psi_{b_{1}}=0,
	\end{equation}
	\begin{equation}
		J_{1}\psi_{a_{n}}+J_{2}\psi_{a_{n+1}}=0,(n=1,2,\cdots),
	\end{equation}
	\begin{equation}
		J_{2}\psi_{b_{n}}+J_{1}\psi_{b_{n+1}}=0,(n=1,2,\cdots),
	\end{equation}
\end{subequations}
The zero-energy edge state is derived as~
\begin{equation}
	|L\rangle=|\psi_{a_{1}},0,\xi\psi_{a_{1}}, 0, \cdots, \xi^{n-1}\psi_{a_{1}},0, \cdots\rangle,
\end{equation}
where $\xi=-J_{1}/J_{2}$ denotes the localization factor. Obviously, the edge state is exponentially localized in the left side of the lattice in topological nontrivial phase and only occupies the \textit{a}-type sites. Similiarly, for a semi-infinite long SSH lattice with right boundry, the zero-energy edge state can be derived as~
\begin{equation}
	|R\rangle=|\cdots, 0,\xi^{N-n}\psi_{b_{N}}, 0, \cdots, 0,\xi\psi_{b_{N}},0,\psi_{b_{N}} \rangle,
\end{equation}
which is exponentially localized in the right side of the lattice in topological nontrivial phase and only occupies the \textit{b}-type sites.

For an even-sized SSH lattice composed of finite sites, we can get its eigenvalues and corresponding eigenstates by diagnolizing the real-space Hamiltonian under base vectors $|L\rangle$ and $|R\rangle$ ~
\begin{equation}
	H_{M}^{\prime}=\left(\begin{array}{cc}
		O_{L,L} & O_{L,R} \\
		O_{R,L} & O_{R,R}
	\end{array}\right),
\end{equation}
where $O_{L,L}=\left\langle L \mid H_{M} \mid L \right\rangle=0$, $O_{L,R}=\left\langle L \mid H_{M} \mid R \right\rangle=\frac{-J_{2}\xi^{N}(\xi^{2}-1)}{\xi^{2 N}-1}$, $O_{R,L}=\left\langle R \mid H_{M} \mid L \right\rangle=O_{L,R}^{*}$, and $O_{R,R}=\left\langle R \mid H_{M} \mid R \right\rangle=0$. The eigenvalues and corresponding eigenstates are~
\begin{subequations}
	\begin{equation}
		E_{0,\pm}=\pm\left|O_{L,R}\right|,
	\end{equation}
	\begin{equation}
		\left|\Psi_{0,\pm}\right\rangle=(|L\rangle\pm|R\rangle) / \sqrt{2},
	\end{equation}
\end{subequations}
Obviously, for an even-sized SSH lattice composed of finite sites, energies of the hybridized edge states in topological nontrivial phase do not equal zero, but a pair of numbers opposite to ecah other due to chiral symmetry~($E_{0,\pm}\rightarrow0$ in the thermodynamic limit of $N \to \infty$). The wavefunctions of almost-zero-energy eigenstates  are odd and even superpositions of states localized exponentially on the left and right edges.

\bibliography{reference}
\end{document}